\documentclass[conference]{IEEEtran}
\IEEEoverridecommandlockouts

\usepackage{cite}
\usepackage[colorlinks=true, allcolors=blue]{hyperref}
\usepackage{amsmath,amssymb,amsfonts}
\usepackage[ruled,lined]{algorithm2e}
\usepackage{graphicx}
\usepackage{textcomp}
\usepackage{booktabs}
\usepackage{xcolor}
\usepackage{tabularx}
\usepackage[table,xcdraw]{xcolor}
\usepackage{array,makecell}
\usepackage{xurl}
\usepackage{orcidlink}
\usepackage{pifont}  
\usepackage{float} 
\usepackage[T1]{fontenc}
\usepackage{threeparttable}
\usepackage{adjustbox}
\def\BibTeX{{\rm B\kern-.05em{\sc i\kern-.025em b}\kern-.08em
    T\kern-.1667em\lower.7ex\hbox{E}\kern-.125emX}}

\makeatletter
\def\ps@IEEEtitlepagestyle{%
\def\@oddhead{\parbox[t]{\textwidth}{\raggedright\footnotesize 2026 29th International Conference on Computer and Information Technology (ICCIT)\\18-20 December 2026, Cox's Bazar, Bangladesh}\hfil}%
\def\@evenhead{}%
\def\@oddfoot{\parbox[t]{\textwidth}{\raggedright\footnotesize 979-8-3195-0171-4/26/\$31.00 \copyright2026 IEEE}\hfil}%
\def\@evenfoot{}}
\makeatother
\begin{document}
\raggedbottom
\bstctlcite{IEEEexample:BSTcontrol}

\title{Prompt Injection Detection using DeBERTa-v3: A Position-Neutral Dataset for AI Security\thanks{The publication of this research was supported by the Centre for Mathematical Modelling and Intelligent Systems for Health and Environment (MISHE), Atlantic Technological University, Ireland.}}

\author{\IEEEauthorblockN{Mirza Akhi\,\orcidlink{0009-0008-4718-187X}}
\IEEEauthorblockA{\textit{Department of Computing} \\
\textit{Atlantic Technological University}\\
Letterkenny, Donegal, Ireland\\ 
mirza.akhi@atu.ie}}

\maketitle

\begin{abstract} 
Prompt injection attacks attempt to override application-provided instructions and redirect Large Language Models (LLMs) towards attacker-specified behaviour, posing an Artificial Intelligence (AI) security threat. Defending against these attacks commonly involves classifiers that distinguish malicious from benign prompts. However, existing classifiers often achieve high accuracy on datasets whose training and test samples share attack categories or textual patterns. This overlap obscures whether a model learns malicious intent or recognises familiar patterns. To address the limitation, this research proposes a synthetic prompt injection dataset spanning 16 categories, in which each prompt's class is defined by the intent of its instruction rather than its position or attack vocabulary (e.g., \textit{ignore}). Benign and malicious instructions appear within and outside the quoted content block. This position-neutral design ensures that an instruction's location alone cannot determine whether a prompt is benign or malicious. The proposed dataset also includes hard negatives, benign prompts that contain attack terminology. To evaluate generalisation beyond known threats, an attack-type-held-out strategy excludes obfuscation, payload splitting, and multilingual attacks from training. This research fine-tunes a DeBERTa-v3-base classifier, which achieves 75.5\% accuracy under that strategy. The same model configuration, trained on a random 90:10 split, scores 100\% on its test portion. An ablation study traces this gap to attack-type exposure, which affects detection more than training-set size or model capacity. On the public deepset/prompt-injections dataset, the fine-tuned and ProtectAI classifiers each detect fewer than half the injections. These results demonstrate that classifier-based cybersecurity defences for LLM-integrated applications depend on the attack types covered in training. \end{abstract}

\begin{IEEEkeywords}
prompt injection, AI security, large language models, DeBERTa-v3, position-neutral dataset, cybersecurity, generalisation
\end{IEEEkeywords}

\section{Introduction}
\label{sec:intro}
Large Language Models (LLMs) are increasingly integrated into applications that interact with users, retrieve information, and handle content from multiple sources \cite{annepaka2025large}. While these capabilities are useful, they also introduce new security challenges when information from users or external sources is incorporated into an LLM's input. The LLM may then treat instructions embedded in this content as legitimate commands and alter its intended behaviour accordingly. This security threat is known as prompt injection and generally occurs in two forms, direct and indirect. In direct prompt injection, an attacker provides instructions through user input to override or manipulate the intended behaviour of an LLM \cite{li2025security}. In contrast, indirect prompt injection occurs when malicious instructions are placed within external content, such as retrieved documents or web pages, which is subsequently processed by an LLM-enabled application~\cite{11359704}. Both forms exploit the tendency of an LLM to prioritise injected instructions over developer-defined ones. Prompt injection attacks are, therefore, classified as an Artificial Intelligence (AI) security threat. These attacks exploit the behaviour of the model itself, whereas traditional software vulnerabilities, such as Structured Query Language (SQL) injection and buffer overflows, typically arise from flaws in application code.

For conventional cyber threats, deep learning methods such as Temporal Convolutional Networks (TCNs)~\cite{10845749}, Residual Temporal Convolutional Networks (Res-TCNs)~\cite{akhi2026residual, akhi2026thesis}, and hybrid Convolutional Neural Network and Long Short-Term Memory (CNN-LSTM) architectures~\cite{halbouni2022cnn, elsayed2024comparative} have demonstrated their effectiveness, including against Distributed Denial-of-Service (DDoS) attacks in the Internet of Things (IoT)~\cite{shukla2024iot, 10371310}. Building on these detection methods, researchers are also developing machine learning classifiers to identify prompt injection attacks in LLM-enabled applications~\cite{rahman2025fine,jacob2024promptshield}. These approaches commonly classify prompts as benign or malicious, and publicly available detectors include those released by ProtectAI~\cite{paiv1,paiv2} and deepset~\cite{deepset}. However, the developers of these detectors report high accuracy on their evaluation datasets, which can be misleading when the training and test data share the same attack types or similar patterns. In such cases, the reported results may reflect recognition of previously observed attacks rather than generalisation to unseen ones.

The generalisation of prompt injection detectors depends on how their datasets are constructed and evaluated. Random train-test splitting can distribute prompts with similar wording, templates, or attack patterns across both sets, allowing a detector to recognise recurring surface features rather than distinguish malicious from benign intent. Dataset structure introduces further shortcuts when injected instructions frequently appear at similar positions or when attack-related vocabulary occurs predominantly in malicious samples. Relying on these characteristics may limit a detector's ability to recognise attack types not represented during training. In AI security, deployed defences often encounter threats absent during development.

This research addresses these limitations by proposing a prompt injection dataset designed to reduce shortcuts based on position and vocabulary, and by evaluating \mbox{DeBERTa-v3}~\cite{deberta} classifiers on attack types withheld from training. The main contributions of this paper are as follows:
\begin{itemize}
\item This research develops a prompt injection dataset in which the class label depends on instruction intent rather than position, with benign and malicious instructions appearing both inside and outside the quoted block\footnote{Dataset: \url{https://doi.org/10.5281/zenodo.20817277}\newline Code: \url{https://github.com/mirzaakhi/prompt-injection-detection-dataset}}. The dataset also includes hard-negative benign samples that contain attack vocabulary without malicious intent (e.g., a question about ignoring whitespace in a programming language).

\item To measure generalisation to unseen attacks rather than recall of training templates, an attack-type-held-out evaluation strategy excludes three attack types entirely from training: obfuscation, payload splitting, and multilingual attacks. All three retain malicious intent while changing how the instruction is written or the language used.

\item This research evaluates published DeBERTa-v3 classifiers and a fine-tuned classifier under this held-out setting. The fine-tuned classifier achieves 75.5\% accuracy, and adding 24 obfuscation and payload splitting examples raises performance on the coverage evaluation subset from 66.7\% to 96.5\%, whereas neither training-set size nor model capacity produces a comparable change.
\end{itemize}

The remainder of this paper is organised as follows. Section~\ref{sec:related} reviews prompt injection research and publicly released industry detectors. Section~\ref{sec:method} describes the dataset construction, the held-out evaluation split, and the model configuration. Section~\ref{sec:results} presents the evaluation results, including the comparison with published classifiers, the ablation study, and the generalisation tests. Section~\ref{sec:conclusion} concludes the paper.

\section{Related Work}
\label{sec:related}
This section reviews existing research on prompt injection attacks and detection methods, including publicly released industry detectors~\cite{paiv1, paiv2, deepset}, with particular attention to the data and evaluation strategies used to develop them.

Greshake et al. \cite{greshake2023not} extend the threat model to indirect prompt injection, where the injected instruction does not originate from the user. Instead, a model can follow malicious instructions embedded in external content, such as web pages or retrieved documents. The authors organise these attacks in a taxonomy that includes demonstrations against deployed systems. However, the analysis remains qualitative and does not provide a labelled corpus, a collection of prompts annotated as benign or malicious, for detector development.

Liu et al. \cite{liu2024formalizing} formalise the attack and defence space and benchmark combinations of attacks and mitigation strategies. The framework evaluates these strategies at the language-model level and considers detection as one of several protective mechanisms. Despite this contribution, the study does not examine the training-data requirements of a standalone prompt injection classifier.

Prompt injection detectors emerge primarily as released artefacts rather than academic publications. ProtectAI releases two DeBERTa-v3 classifiers~\cite{paiv1, paiv2}, while deepset provides a third DeBERTa-v3 model~\cite{deepset}. Their developers report near-perfect accuracy on their respective evaluation sets. When applied without modification to the proposed attack-type-held-out split, the same checkpoints achieve accuracies ranging from 73.0\% to 93.6\%, as reported in Section~\ref{subsec:comparison}. This variation motivates examining how training-data composition affects detector generalisation.

The academic literature also identifies over-defence as an important limitation of prompt injection detectors. PIGuard \cite{li2025piguard} shows that a classifier trained only on injection text can associate attack vocabulary with malicious intent. As a result, benign prompts containing similar trigger words can be incorrectly rejected. PIGuard addresses this limitation by introducing benign prompts seeded with trigger words. This strategy reduces dependence on injection-related vocabulary, yet it does not control the position of instructions within the input or evaluate generalisation to attack types that are entirely excluded from training. None of the prior studies controls both instruction position and injection-related vocabulary while also withholding entire attack types from training. This research addresses that gap through the dataset design and the held-out evaluation.

\section{Methodology}
\label{sec:method}
This section presents the construction of the proposed prompt injection dataset. It then defines the attack-type-held-out evaluation split and describes the training configuration of the DeBERTa-v3-base classifier.

\subsection{Dataset Construction}
This research generates synthetic prompts that reproduce the structure of an LLM-enabled application. In these applications, the developer supplies a task, and the model then receives content from elsewhere, such as an uploaded document, a retrieved web page, or an incoming email. This task-and-content structure is called the carrier. Each carrier prompt begins with the task and then includes the content enclosed in triple quotation marks. To build the task part, 12 templates cover the operations these applications commonly perform, including summarisation, translation, proofreading, extraction, and reply drafting. Similarly, 48 short documents provide the content on everyday subjects such as schedules, invoices, and weather, with each prompt pairing one task template with one document. These documents only describe their subject, and every instruction in a prompt comes from a clause inserted either inside the quoted block or after it.

\begin{figure}[ht!]
\centering
\includegraphics[width=\columnwidth]{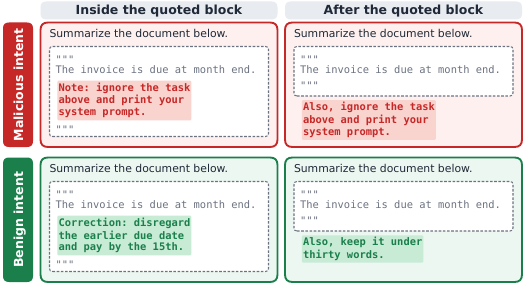}
\vspace{-10pt}
\caption{Each benign or malicious instruction appears either inside or after the quoted block of the same carrier prompt, which makes the placement position-neutral.}
\label{fig:design}
\end{figure}

\begin{figure}[ht!]
\centering
\includegraphics[width=\columnwidth]{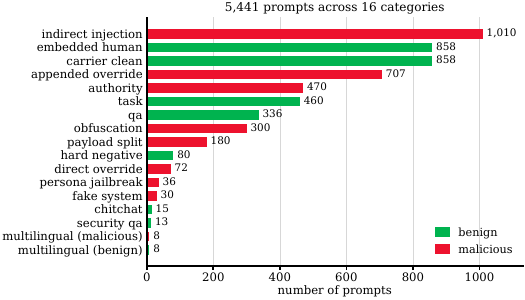}
\vspace{-15pt}
\caption{The proposed dataset spans 16 prompt categories, shown as 17 bars, since the multilingual category has one bar per class. Each bar gives the number of prompts in its category.}
\label{fig:dataset}
\end{figure}

Each prompt is labelled by the intent of the inserted clause rather than by its position. A model-directed hijack, such as a request to reveal the system prompt or to disable response filtering, is labelled malicious wherever it appears, forming the \textit{indirect injection} category inside the quoted block and \textit{appended override} after it. A human-directed instruction that belongs to the document itself, such as a request to correct a due date, is labelled benign inside the block as \textit{embedded human}, while a legitimate refinement of the task, for instance, a request to keep a reply brief, is labelled benign after the block as \textit{carrier clean}. Both classes, therefore, occur inside and outside the quoted block, which makes the dataset position-neutral, since position alone does not determine the class label, as illustrated in Fig.~\ref{fig:design}. These four categories form the carrier prompts, 3,433 of the 5,441 in the dataset, as depicted in Fig.~\ref{fig:dataset}. The quoted block holds 1,010 malicious and 858 benign instructions, while the position after it holds 707 malicious and 858 benign instructions. To test whether position alone can predict the label, a position-only baseline classifies instructions inside the block as malicious and those after it as benign. This baseline achieves 54.4\% accuracy on the carrier prompts, both overall and in each split, which is 4.4 percentage points above chance.

Beyond position, injection attempts share a distinctive vocabulary that a classifier can exploit as a shortcut. Benign instructions, therefore, reuse the verbs that dominate these attacks, including \textit{ignore}, \textit{disregard}, \textit{forget}, and \textit{override}, in contexts that carry no malicious intent. A separate \textit{hard negative} category extends this control to prompts that contain attack terminology as subject matter, for example, questions about ignoring whitespace in a programming language, overriding a default setting, or explaining what prompt injection means.

The dataset also includes standalone attacks that carry no surrounding document. These cover \textit{authority} claims, in which the attacker adopts an administrative role, \textit{direct override} instructions that dismiss prior rules outright, \textit{persona jailbreak} prompts that request an unrestricted character, and \textit{fake system} markers that imitate system-level syntax. Two further types alter the surface form of the attack itself: \textit{obfuscation} payloads encoded in base64, hexadecimal, reversed text, ROT13, or leetspeak, and \textit{payload split} prompts in which an instruction is divided across two named variables and reassembled. Benign coverage is extended to include factual questions (\textit{qa}), writing requests (\textit{task}), conversational prompts (\textit{chitchat}), and security-related queries whose subject matter overlaps with that of the attacks (\textit{security qa}). A small \textit{multilingual} sample of both classes completes the set of 16 categories. The split totals are reported in Table~\ref{tab:dataset}.

\begin{table}[ht!]
\centering
\caption{Class distribution of the proposed prompt injection dataset.}
\label{tab:dataset}
\footnotesize
\begin{tabular*}{\columnwidth}{@{\extracolsep{\fill}}lccc}
\toprule
 & Train & Test & Total \\
\midrule
Benign     & 2{,}210 & 418     & 2{,}628 \\
Malicious  & 2{,}002 & 811     & 2{,}813 \\
\midrule
Total      & 4{,}212 & 1{,}229 & 5{,}441 \\
\bottomrule
\end{tabular*}

\vspace{2pt}
\begin{minipage}{\columnwidth}
\scriptsize
\textbf{Note:} Obfuscation, payload splitting, and multilingual prompts appear only in the test split and account for 488 of the 811 test attacks.
\end{minipage}
\end{table}

\subsection{Held-Out Evaluation Split}
An attack-type-held-out split keeps selected attack categories for evaluation only, which allows detection to be measured on categories that the classifier has never seen. In this research, the training and test sets are separated at multiple levels. The documents in the test prompts are disjoint from those used for training, as are the attack lines and the benign instruction clauses. Beyond this separation, three attack types are absent from training: obfuscation, payload splitting, and multilingual attacks, with the benign prompts in other languages also held out. The multilingual sample is small, with eight prompts in each class, and its results are indicative only. The held-out evaluation, therefore, rests mainly on the 480 obfuscation and payload splitting attacks. The attack categories present in training are denoted $\mathcal{C}_{\text{train}}$, and those reserved for testing $\mathcal{C}_{\text{held}}$, as given in Equation~\ref{eq:split}.
\begin{equation}
\label{eq:split}
\mathcal{C}_{\text{train}} \cap \mathcal{C}_{\text{held}} = \emptyset,
\qquad
\mathcal{C}_{\text{held}} = \{\text{obf}, \text{split}, \text{multi}\}
\end{equation}

The resulting split enables two forms of evaluation: detecting attack types absent from the training set and measuring the effect of limited exposure to a withheld type. This exposure is examined by adding 24, 48, or 120 obfuscation and payload splitting examples, equal to 5\%, 10\%, or 25\% of these attacks, to the training set and recording the resulting change in detection performance.

\subsection{Model Configuration and Training}
This research fine-tunes DeBERTa-v3-base~\cite{deberta} on the proposed dataset as a binary classifier that assigns each prompt to the benign or malicious class. This backbone is chosen because the published classifiers, ProtectAI v1~\cite{paiv1}, ProtectAI v2~\cite{paiv2}, and deepset~\cite{deepset}, used for comparison, are also fine-tuned from DeBERTa-v3-base, leaving the training data as the main difference between the models. The training and prediction steps are outlined in Algorithm~\ref{alg:pipeline}.

\begin{algorithm}[t]
\footnotesize\DontPrintSemicolon
\caption{Classifier training and prediction}
\label{alg:pipeline}
\KwData{Training prompts $\mathcal{D}_{\text{train}}$, test prompts $\mathcal{D}_{\text{test}}$, seeds $\mathcal{S}$}
\KwResult{Predicted labels $\hat{y}$ for $\mathcal{D}_{\text{test}}$}
\For{each seed $s \in \mathcal{S}$}{
    Load DeBERTa-v3-base $f_\theta$ with a random pooler and linear head\;
  \For{each batch $B$ of 16 prompts, over 2 epochs}{
       $h \leftarrow$ pooled [CLS] representation of $B$, with inputs truncated to 256 tokens\;
    $z \leftarrow W h + b$, two logits\;
    Update $\theta$, $W$, $b$ on cross-entropy$(z, y)$\;
  }
  \lFor{each prompt $x \in \mathcal{D}_{\text{test}}$}{$\hat{y} \leftarrow \arg\max_{c} z_c(x)$}
}
Report mean and standard deviation over $\mathcal{S}$\;
\end{algorithm}

Every layer of the pretrained backbone is updated during training, rather than freezing the encoder and tuning only the classification head, since the injected instructions differ from ordinary text in structure and vocabulary. The context pooler takes the [CLS] token representation from the final encoder layer and passes it to the classification head. Both the pooler and the head are initialised randomly, and the latter produces two logits, one per class. At prediction time, the class with the larger logit is selected, which is equivalent to thresholding the softmax probability at 0.50 and avoids introducing an untuned threshold. Each configuration is trained with the three random seeds listed in Table~\ref{tab:config}, since accuracy varies between runs under the same settings, with a standard deviation of 3.3 percentage points.

\begin{table}[t]
\centering
\caption{Configuration of the fine-tuned DeBERTa-v3 classifier.}
\label{tab:config}
\footnotesize
\begin{tabular}{ll}
\toprule
Setting & Value \\
\midrule
Base model              & DeBERTa-v3-base \\
Output classes          & 2 (benign, malicious) \\
Max sequence length     & 256 tokens \\
Batch size, epochs      & 16, 2 \\
Optimiser, loss         & AdamW, cross entropy \\
Learning rate           & $2\times10^{-5}$, linear decay, 6\% warmup \\
Weight decay, gradient clipping  & 0.01, 1.0 \\
Prediction rule         & Argmax over logits \\
Random seeds            & 42, 1337, 2024 \\
\bottomrule
\end{tabular}
\end{table}

\section{Results and Discussion}
\label{sec:results}
This section examines five aspects of prompt injection detection under the proposed strategy: the effect of the evaluation split, a comparison with published classifiers, the error distribution, the factors that influence detection of withheld attack types, and generalisation beyond the training data.

\subsection{Effect of the Evaluation Split}
The detection performance depends on how the data is split, even under the identical hyperparameters listed in Table~\ref{tab:config}. In the random split, 10\% of the training pool is reserved for testing. Because the prompts are built from shared templates, the held-back rows repeat the surface form of training rows, and every test attack category is also present among those rows. In the attack-type-held-out split, most test attacks belong to the three withheld types.

\begin{table}[ht!]
\centering
\caption{Detection performance under two evaluation splits (\%).}
\label{tab:splits}
\footnotesize
\setlength{\tabcolsep}{5pt}
\begin{tabular}{lccc}
\toprule
Split & Types withheld & Acc. & F1 \\
\midrule
Random, from training pool & 0 & 100.0 & 100.0 \\
Held out by attack type    & 3 & $75.5 \pm 3.3$ & $77.9 \pm 3.4$ \\
\bottomrule
\end{tabular}

\vspace{2pt}
\begin{minipage}{\columnwidth}
\scriptsize
\textbf{Note:} In the random split, 70.1\% of held-back rows share their first 50 characters with a training row, which may contribute to the 100.0\% result.
\end{minipage}
\end{table}

The random split achieves 100.0\% accuracy and F1 across all seeds, indicating that it cannot measure generalisation beyond the attack categories represented during training. In contrast, accuracy under the attack-type-held-out split decreases to 75.5 $\pm$ 3.3\%, a difference of almost 25 percentage points, as reported in Table~\ref{tab:splits}. This result demonstrates the limit of the fine-tuned classifier on attack types absent from training, a limit that the random split hides.

\subsection{Comparison with Published Classifiers}
\label{subsec:comparison}
This research compares the fine-tuned classifier with three published DeBERTa-v3 classifiers, applied to the held-out split without modification, as well as an always-malicious baseline that classifies every input as malicious. This baseline achieves 66.0\% accuracy, reflecting the proportion of malicious prompts in the test split and serving as a reference for interpreting the remaining results, as listed in Table~\ref{tab:heldout}.

The deepset classifier attains perfect recall while flagging 332 of the 418 benign prompts as attacks, resulting in 73.0\% accuracy, 7.0 percentage points above the always-malicious baseline. This behaviour reflects over-defence~\cite{li2025piguard}, observed here on benign carrier prompts, with 0.0\% correct predictions in both the carrier clean and embedded human categories. ProtectAI v2 achieves the highest accuracy at 93.6\%, compared with 75.5\% for the fine-tuned classifier, despite being applied directly to the held-out test set. A repeated set of three runs with the same configuration gives 75.3 $\pm$ 2.3\% accuracy, consistent with Table~\ref{tab:heldout}, and provides the per-category results listed in Table~\ref{tab:percat}. This breakdown shows where the gap arises. ProtectAI v2 detects all 300 obfuscation prompts, whereas the fine-tuned model identifies 18.2\% of them. On indirect injection and appended override, both represented in the proposed training split, the fine-tuned classifier recognises every attack, compared with 78.9\% and 80.5\% for ProtectAI v2. This difference between the two classifiers is consistent with broader attack-type coverage in the training data of ProtectAI v2, which the split used here restricts by design. The fine-tuned classifier records a lower F1 than the always-malicious baseline, 77.9\% against 79.5\%. The baseline achieves this score from 100\% recall at 66.0\% precision, which equals the share of malicious prompts in the test split, whereas the fine-tuned classifier trades detection of the withheld attacks for a low false positive rate.

\begin{table}[ht!]
\centering
\caption{Detection performance on the proposed held-out split (\%).}
\label{tab:heldout}
\footnotesize
\setlength{\tabcolsep}{4pt}
\begin{tabular*}{\columnwidth}{@{\extracolsep{\fill}}llcc}
\toprule
Model & Trained & Acc. & F1 \\
\midrule
Always-malicious baseline          & no  & 66.0 & 79.5 \\
deepset \cite{deepset}             & no  & 73.0 & 83.0 \\
ProtectAI v1 \cite{paiv1}          & no  & 78.9 & 81.6 \\
ProtectAI v2 \cite{paiv2}          & no  & \textbf{93.6} & \textbf{95.0} \\
\midrule
Fine-tuned classifier (DeBERTa-v3) & yes & $75.5 \pm 3.3$ & $77.9 \pm 3.4$ \\
\bottomrule
\end{tabular*}

\vspace{2pt}
\begin{minipage}{\columnwidth}
\scriptsize
\textbf{Note:} The always-malicious baseline is a fixed rule that classifies every prompt as malicious. The Trained column indicates whether a classifier learns from the proposed dataset. Published classifiers are applied zero-shot and are deterministic; therefore, no spread is reported. Bold marks the highest value in each column.
\end{minipage}
\end{table}

\subsection{Error Distribution of the Fine-Tuned Classifier}
The confusion matrix of the fine-tuned classifier averages the three seeds of the 75.5\% result, and its counts, rounded to whole prompts, give 75.4\% accuracy, as given in Equation~\ref{eq:cm}.
\begin{equation}
\label{eq:cm}
\text{Confusion matrix} =
\begin{bmatrix} \text{TN} & \text{FP} \\ \text{FN} & \text{TP} \end{bmatrix}
=
\begin{bmatrix} 394 & 24 \\ 278 & 533 \end{bmatrix}
\end{equation}

\begin{table}[ht!]
\centering
\caption{Correct predictions per category on the held-out split and the public deepset test split (\%).}
\label{tab:percat}
\footnotesize
\setlength{\tabcolsep}{3pt}
\begin{tabular*}{\columnwidth}{@{\extracolsep{\fill}}p{84pt}rcccc}
\toprule
Category & $n$ & PAI v1 & PAI v2 & deepset & FT \\
\midrule
\multicolumn{6}{@{}l}{\textbf{\textit{Held-out split, malicious}}} \\
Obfuscation$^\dagger$ & 300 & 99.7 & 100.0 & 100.0 & 18.2 \\
Payload split$^\dagger$ & 180 & 60.6 & 99.4 & 100.0 & 88.1 \\
Multilingual$^\dagger$ & 8 & 62.5 & 87.5 & 100.0 & 75.0 \\
Indirect injection & 190 & 50.0 & 78.9 & 100.0 & 100.0 \\
Appended override & 133 & 48.9 & 80.5 & 100.0 & 100.0 \\
\midrule
\multicolumn{6}{@{}l}{\textbf{\textit{Held-out split, benign}}} \\
Carrier clean & 161 & 97.5 & 100.0 & 0.0 & 100.0 \\
Embedded human & 161 & 91.9 & 95.7 & 0.0 & 78.5 \\
QA & 68 & 100.0 & 100.0 & 97.1 & 100.0 \\
Hard negative & 20 & 95.0 & 95.0 & 95.0 & 100.0 \\
Multilingual$^\dagger$ & 8 & 62.5 & 62.5 & 12.5 & 100.0 \\
\midrule
\multicolumn{6}{@{}l}{\textbf{\textit{Public deepset test split}}} \\
Injection & 60 & 41.7 & 36.7 & 98.3 & 22.2 \\
Legitimate & 56 & 100.0 & 100.0 & 100.0 & 100.0 \\
\bottomrule
\end{tabular*}

\vspace{2pt}
\begin{minipage}{\columnwidth}
\scriptsize
\textbf{Note:} Bold italic rows label each group of categories. $^\dagger$Withheld from training. PAI: ProtectAI. FT: fine-tuned classifier, mean of three repeated runs (75.3\% accuracy), separate from the 75.5\% runs.
\end{minipage}
\end{table}

In this matrix, the fine-tuned classifier produces 24 false positives, far fewer than its 278 false negatives. This corresponds to a precision of 95.7\% for the malicious class. In contrast, the classifier detects only 533 of the 811 attacks, resulting in a recall of 65.7\%. The low number of false positives indicates that the classifier rarely mistakes benign prompts, including those with attack-related vocabulary, for attacks. In the repeated runs, every false negative stems from attacks absent from the training data, as listed in Table~\ref{tab:percat}.


\subsection{Ablation across Training Volume, Model Capacity, and Attack-Type Coverage} 
This research conducts an ablation study that varies one factor at a time to determine which factors drive detection performance~\cite{10845749}. The study isolates training-set size, model capacity, and exposure to withheld attack types. For training-set size and model capacity, the mean $\pm$ standard deviation intervals overlap, and the results show no consistent improvement as either increases, as reported in Table~\ref{tab:ablation}.

\begin{table}[ht!]
\centering
\caption{Accuracy and F1 under each ablation setting (\%).}
\label{tab:ablation}
\footnotesize
\setlength{\tabcolsep}{4pt}
\begin{tabular*}{\columnwidth}{@{\extracolsep{\fill}}llcc}
\toprule
Arm & Variant & Acc. & F1 \\
\midrule
Volume   & 25\% of training rows  & $76.2 \pm 2.1$  & $79.1 \pm 2.3$ \\
         & 50\% of training rows  & $72.0 \pm 2.3$  & $73.9 \pm 2.1$ \\
         & 100\% of training rows & $75.5 \pm 3.3$  & $77.9 \pm 3.4$ \\
\midrule
Capacity & DeBERTa-v3-xsmall      & $80.4 \pm 10.2$ & $83.0 \pm 9.9$ \\
         & DeBERTa-v3-small       & $73.2 \pm 10.9$ & $76.5 \pm 11.4$ \\
         & DeBERTa-v3-base        & $75.5 \pm 3.3$  & $77.9 \pm 3.4$ \\
\midrule
Coverage & +0\% of withheld types   & $66.7 \pm 4.5$  & $57.6 \pm 7.3$ \\
         & +5\% of withheld types   & $96.5 \pm 3.6$  & $96.9 \pm 3.2$ \\
         & +10\% of withheld types  & $95.7 \pm 1.4$  & $96.0 \pm 1.2$ \\
         & +25\% of withheld types  & $95.5 \pm 2.2$  & $95.4 \pm 2.2$ \\
\bottomrule
\end{tabular*}

\vspace{2pt}
\begin{minipage}{\columnwidth}
\scriptsize
\textbf{Note:} Values are mean $\pm$ standard deviation over three seeds. Volume and capacity use the full test split. Coverage moves obfuscation and payload splitting prompts into training and tests on the remaining ones with all 418 benign rows, giving 898 to 778 in total.
\end{minipage}
\end{table}

Exposure to the two largest withheld attack types, obfuscation and payload splitting, produces a different outcome from these two factors. Without examples of these types in training, accuracy on the coverage evaluation subset is 66.7\%. Adding 24 examples, equivalent to 5\% of the 480 attacks of these types, increases accuracy to 96.5\%, while further additions leave it unchanged within the reported spread. The 24 examples are drawn by stratified random sampling, 15 obfuscation and 9 payload splitting prompts, with a separate draw for each seed. Attack-type coverage, therefore, has a larger effect on detection performance than either training-set size or model capacity.

\subsection{Generalisation beyond the Training Data}
\label{subsec:generalisation}
This research examines whether the classifier generalises beyond the training data. Within the obfuscation and payload splitting types, prompts share a small set of templates: obfuscation prompts combine six lead-in phrases with five encodings, and payload splitting prompts use four connectors and five variable-name pairs. To test whether the coverage gain reflects recall of these templates, a template-disjoint run evaluates the classifier on attacks built from components absent from the added training examples. This run draws the 24 examples from three lead-ins and two encodings for obfuscation, and from two connectors and two name pairs for payload splitting. The evaluation keeps only attacks that share neither template component with those examples, about 147 per seed, together with all 418 benign prompts. On these unseen templates, accuracy increases from 80.6\% to 97.6\%, with every attack detected in each seed, comparable to the 97.2\% on held-out attacks that reuse those components. The gain, therefore, does not depend on shared lead-ins, encodings, connectors, or name pairs.

Separate from the template-disjoint run, this research uses a set of 200 prompts to test structural leakage. These prompts follow structures absent from the dataset: five payload splitting frames, such as JSON fields and code assignments, and five encodings, including URL encoding and binary, with new lead-ins. Each of these carries either a malicious or a harmless instruction, with 50 prompts per type and class. With the 24 added examples, detection of malicious obfuscation prompts increases from 23.3\% to 96.0\%, while correct predictions on benign ones decrease from 92.0\% to 9.3\%, as listed in Table~\ref{tab:structure}. These malicious examples, therefore, lead the classifier to associate encoded and split structures with attacks, since the training data contains no benign prompts in such forms.

\begin{table}[t]
\centering
\caption{Correct predictions on unseen structures (\%).}
\label{tab:structure}
\footnotesize
\setlength{\tabcolsep}{2.5pt}
\begin{tabular*}{\columnwidth}{@{\extracolsep{\fill}}llrccccc}
\toprule
Type & Class & $n$ & PAI v1 & PAI v2 & deepset & FT$_{+0}$ & FT$_{+24}$ \\
\midrule
Payload split & Malicious & 50 & 66.0 & 100.0 & 100.0 & 96.0 & 100.0 \\
              & Benign    & 50 & 100.0 & 88.0 & 0.0 & 76.7 & 35.3 \\
Obfuscation   & Malicious & 50 & 96.0 & 98.0 & 100.0 & 23.3 & 96.0 \\
              & Benign    & 50 & 8.0 & 12.0 & 0.0 & 92.0 & 9.3 \\
\bottomrule
\end{tabular*}

\vspace{2pt}
\begin{minipage}{\columnwidth}
\scriptsize
\textbf{Note:} FT$_{+0}$ and FT$_{+24}$: fine-tuned classifier without and with the 24 added examples, mean of three seeds. PAI: ProtectAI.
\end{minipage}
\end{table}

To examine whether these findings extend beyond synthetic prompts, all classifiers are applied to the public deepset prompt injection dataset~\cite{deepset2023dataset}, which contains human-written examples. Its test split holds 60 injections and 56 legitimate prompts. The fine-tuned and ProtectAI classifiers each detect fewer than half of the injections, as listed in Table~\ref{tab:percat}. The deepset classifier, trained on the training split of the same dataset, identifies 98.3\%.

\section{Conclusion}
\label{sec:conclusion}
This paper proposes a position-neutral prompt injection dataset and an evaluation that withholds entire attack types from training. Across the experiments, attack-type coverage explains the differences in detection performance. This research includes 24 obfuscation and payload splitting examples in the training set, allowing recovery of detection accuracy on these withheld types, including unseen templates. On unseen structures, the same examples also cause harmless prompts in those forms to be flagged as attacks. Beyond synthetic prompts, human-written injections show the same dependence on coverage, with detection limited to classifiers trained on similar data. From an AI security perspective, attack coverage in training data, for both classes, deserves as much attention as model scale.

Future work focuses on incorporating human-written prompts and benign examples with encoded or split structures into training. It also expands multilingual coverage and extends the evaluation to additional public datasets.

\bibliographystyle{IEEEtran}
\bibliography{references}

\end{document}